\documentclass{lhep}    
\usepackage{amssymb,amsmath,amsbsy}
\usepackage{dsfont}
\usepackage{mathrsfs}
\newenvironment{Eqnarray}%
     {\arraycolsep 0.14em\begin{eqnarray}}{\end{eqnarray}}
\def\beq{\begin{equation}}
\def\eeq{\end{equation}}
\def\beqa{\begin{Eqnarray}}
\def\eeqa{\end{Eqnarray}}

\def\half{\tfrac12}
\def\rsub#1{_{\raise 1.5pt\hbox{$\scriptstyle#1$}}}
\def\lsup#1{^{\lower 6pt\hbox{$\scriptstyle#1$}}}
\def\llsup#1{^{\lower 3pt\hbox{$\scriptstyle#1$}}}
\def\lowsup#1{^{\lower 2pt\hbox{$\scriptstyle#1$}}}
\def\lsim{\mathrel{\raise.3ex\hbox{$<$\kern-.75em\lower1ex\hbox{$\sim$}}}}
\def\gsim{\mathrel{\raise.3ex\hbox{$>$\kern-.75em\lower1ex\hbox{$\sim$}}}}

\def\msusyy{M_{\rm S}^2}
\def\nn{\nonumber}
\def\cbma{\cos(\beta-\alpha)}
\def\sbma{\sin(\beta-\alpha)}

\journal{LHEP}
\vol{xx}
\jyear{2023}
\pages{xxx} 

\received{xx January 2018}
\published{xx March 2018}

\begin{document}

\title{Higgs Boson Physics---The View Ahead}

\author{Howard E.~Haber}
\address{Santa Cruz Institute for Particle Physics, University of California, Santa Cruz, CA 95064, USA}

\begin{abstract}
Eleven years ago, the Higgs boson was discovered at the LHC.   I briefly survey the status of Higgs boson physics today and explore some of the implications for future Higgs studies.   Although current experimental measurements are consistent with interpreting the observed Higgs boson as being consistent with the predictions of the Standard Model of particle physics, it is still possible that the Higgs boson is a member of an extended scalar sector that lies beyond the Standard Model.  Nevertheless, an extended Higgs sector is already highly constrained.   The Higgs sector can also serve as a portal to new physics beyond the Standard Model.   Finally, two Higgs wishlists are assembled that merit future study and clarification at the LHC and future collider facilities now under development.
\end{abstract}

\maketitle


\section{Introduction}
\label{intro}

The $W^\pm$ and $Z$ gauge bosons, which mediate the weak nuclear force, were discovered in 1983 (see, e.g., L.~DiLella and C.~Rubbia in ~\cite{Schopper:2015njg}).   
Due to electroweak symmetry breaking, nonzero masses for the $W^\pm$ and $Z$ are generated, whereas
the photon (which mediates the electromagnetic force) remains massless.
A priori, one could have imagined a plethora of possible dynamical models responsible for electroweak symmetry breaking.  Remarkably, the simplest possible model, a self-interacting complex doublet of elementary scalar fields that yields a physical scalar particle---the Higgs boson, is consistent with all current experimental data.  Nevertheless, there are a number of profound theoretical questions which suggest that the complete story of electroweak symmetry breaking has not yet been written.
In this short review, I will briefly discuss the current status of Higgs boson physics and explore some of the implications for future Higgs studies.

\section{The current status of the Higgs boson}
\label{status}

\begin{figure}[t!]
\centering
\hspace{-0.15in}
    \includegraphics[width=0.5\textwidth]{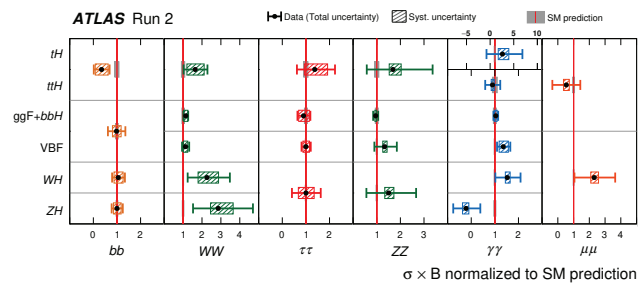} 
	\caption{Ratio of observed rate to predicted SM event rate for different combinations of Higgs boson production and decay processes, as observed by the ATLAS Collaboration (based on 139 fb$^{-1}$ of data). 
	The horizontal bar on each point denotes the 68\% confidence interval. The narrow grey bands indicate the theory uncertainties in the SM cross section times the branching fraction predictions. The $p$-value for compatibility of the measurement and the SM prediction is 72\%.  
	Taken from~\cite{ATLAS:2022vkf}. \label{ATLAS-Higgs}}	
\end{figure}

On July 4, 2012, the discovery of an electrically neutral scalar particle with a mass of about 125 GeV at the Large Hadron Collider (LHC) was announced (see, e.g., P.~Jenni and T.S. Virdee in ~\cite{Schopper:2015njg}).   After ten years of data collection and analysis, it has been determined that the properties of this scalar particle closely resemble those of the Higgs boson that was predicted by the Standard Model (SM) of particle physics~\cite{Langacker:2017uah}.  At the LHC, Higgs bosons are produced by a variety of mechanisms, the most important of which is gluon-gluon fusion (ggF).   Once produced, the Higgs boson (denoted by $H$) can decay into a variety of final states.   The relevant experimental observable is the product of the cross section ($\sigma$) of a given production mechanism multiplied by the branching ratio ($B$) for Higgs boson decay into the observed final state.
In Figure ~\ref{ATLAS-Higgs}, the product $\sigma B$ for 
the various Higgs production mechanisms and final state decay products, observed by the ATLAS Collaboration~\cite{ATLAS:2022vkf}, is compared with the corresponding SM predictions.  A similar plot obtained by the CMS Collaboration can be found in \cite{CMS:2022dwd}.

\begin{figure}[t!]
\centering
    \includegraphics[width=0.43\textwidth]{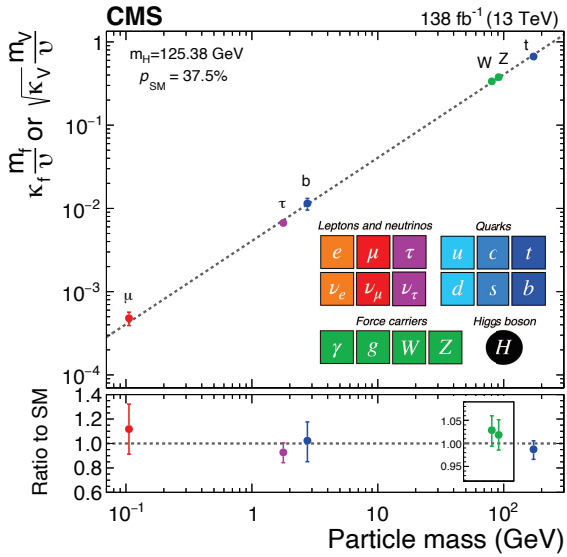} 
	\caption{The measured coupling modifiers of the Higgs boson to fermions and heavy gauge bosons, observed by the CMS Collaboration, as functions of fermion or gauge boson mass, where $v$ is the vacuum expectation value of the Higgs field. For gauge bosons, the square root of the coupling modifier is plotted, to keep a linear proportionality to the mass, as predicted in the SM. The $p$-value with respect to the SM prediction is 37.5\%.   Taken from \cite{CMS:2022dwd}.\label{CMS-Higgs}\\[-11pt]} 
\end{figure}

The Higgs boson is intimately connected with the mechanism of mass generation for the fundamental particles of the SM (quarks, charged leptons, and gauge bosons).
As a result, the SM Higgs boson couples to fermions [gauge bosons] with a strength that is proportional to the corresponding fermion mass [gauge boson squared mass].  This expected behavior is experimentally reproduced in Figure \ref{CMS-Higgs} based on results obtained by the CMS Collaboration~\cite{CMS:2022dwd}.   A similar plot obtained by the ATLAS Collaboration can be found in \cite{ATLAS:2022vkf}.  The results shown in Figures \ref{ATLAS-Higgs} and \ref{CMS-Higgs} provide a quantitative measure of the consistency of the Higgs boson properties with those predicted by the SM.

Although the SM has been remarkably successful in describing the behavior of the fundamental particles and their interactions, the SM cannot be the ultimate theory of particle physics.
Putting aside a few experimental anomalies that could be statistical flukes or could be evidence of cracks in the SM, there exist a number of important clues that strongly suggest the need 
for new physical laws beyond the Standard Model (BSM).   For example, the 
astrophysical evidence for dark matter cannot be explained in the context of the SM, and may be a consequence of a new particle associated with BSM physics~\cite{Profumo:2017hqp}.  Another hint for BSM physics arises when attempting to devise a theory of baryogenesis in the early universe to explain the observed asymmetry between baryons and their antiparticles, which cannot be explained by SM physics alone~\cite{White:2016nbo}.  Finally, the origin of the energy scale of electroweak symmetry breaking remains a mystery.  
Employing an old argument first advanced by Weisskopf in 1939~\cite{Weisskopf:1939zz}, the quadratic divergence of the self-energy of an elementary scalar $H$ is interpreted as implying the existence of a new energy scale of order $\Lambda\sim m_H/g\sim 1$~TeV (where $g$ is a weak coupling constant).  
The scale $\Lambda$ characterizes the BSM physics that governs the fundamental dynamics of electroweak symmetry breaking.   The TeV energy scale is now being probed at the LHC, although so far no evidence for BSM physics has been detected.  In the absence of new physics near the TeV scale, the Higgs sector of the SM is \textit{unnatural}; see \cite{Craig:2022uua} for further details.

\begin{figure}[t!]
\centering
    \includegraphics[width=0.4\textwidth]{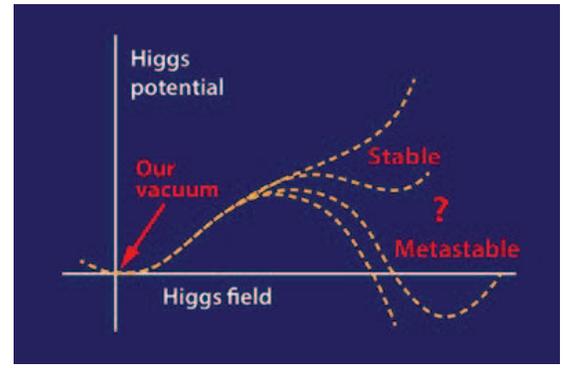} 
	\caption{The electroweak vacuum may not be stable depending on the behavior of the scalar potential at large Higgs field values (figure courtesy of A.~Kusenko). 
	\label{metastable}}
\end{figure}

In many approaches, hints for the existence of BSM physics will emerge from theoretical and experimental studies of the Higgs sector.   Here is one interesting example.  The Higgs field of the SM has a local minimum at a field value of $v=246$~GeV. However, it is possible that a second minimum develops at very large scalar field values as shown in Figure \ref{metastable}. 
For field values larger than the Planck energy scale, $M_{\rm P}= 10^{19}$~GeV, calculations within the SM are not reliable, as gravitational effects can no longer be neglected.  However, at field values below $M_P$, one can reliably compute the shape of the SM scalar potential to determine whether our vacuum is stable.   
%
\begin{figure}[h!]
\centering
    \includegraphics[width=0.45\textwidth]{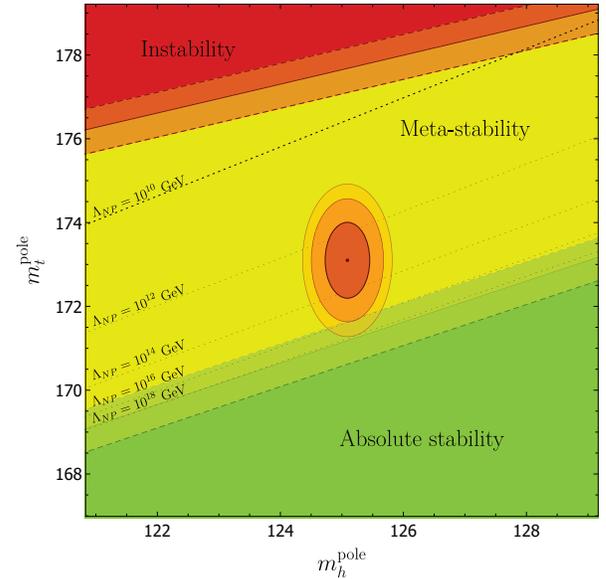}
	\caption{The results of a computation presented in \cite{Andreassen:2017rzq} indicate that the electroweak vacuum of the SM is metastable.  The ellipses show the 68\%, 95\%, and 99\% contours based on the experimental uncertainties in the top quark and Higgs boson pole masses.  The dotted lines indicate the energy scale at which the addition of higher-dimensional operators could stabilize the SM.
	\label{metastable2}}
\end{figure}

As shown in Figure \ref{metastable2}, a theoretical computation most recently carried out in \cite{Andreassen:2017rzq} indicates that in the SM, the electroweak vacuum is in fact metastable (after using the observed values of the top quark and Higgs mass along with the measured value of the strong coupling constant $\alpha_s$ evaluated at the $Z$ boson mass).  The same authors estimate that the lifetime of this metastable vacuum is $10^p$~years, where $p=139^{+102}_{-51}$.  That is, the lifetime of the electroweak vacuum is enormous as compared with the age of the universe (which is roughly 14 billion years).   Although this result alleviates any concern about the future demise of our universe, it is nevertheless unsettling that our present vacuum may not be absolutely stable. Moreover, as indicated in Figure~\ref{metastable2}, if new BSM physics enters at an energy scale between LHC energies and the Planck scale, then the metastability properties of the electroweak vacuum could be significantly altered (and perhaps rendered absolutely stable).

\begin{figure}[t!]
\centering
    \includegraphics[width=0.4\textwidth]{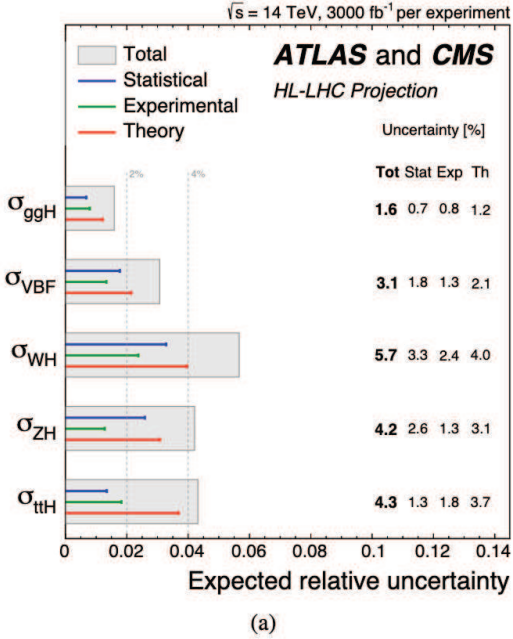}  \\
     \includegraphics[width=0.4\textwidth]{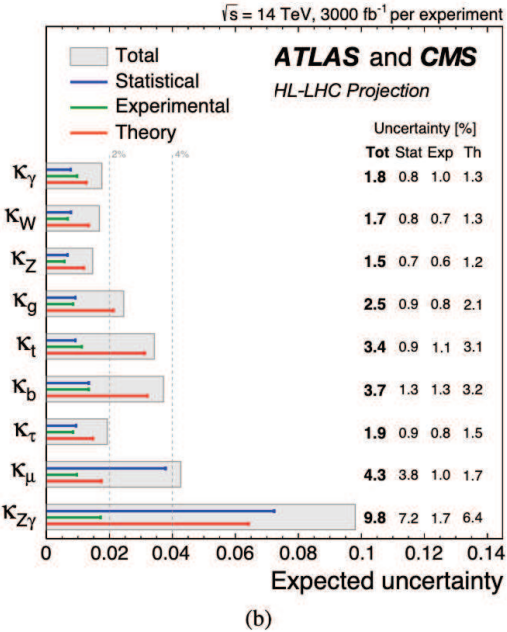} 
	\caption{Summary plots showing the total expected uncertainties on (a) the per-production-mode cross-sections
normalized to the SM predictions and (b) the coupling modifier parameters ($\kappa$), for the combination of ATLAS and
CMS extrapolations. For each measurement, the total uncertainty is indicated by a grey box while the statistical,
experimental, and theory uncertainties are indicated by a blue, green, and red line, respectively.  Taken from \cite{ATLAS:2022hsp}.\label{HLLHC1}}
\end{figure}

\begin{figure}[t!]
   \hspace{-0.4cm}
    \includegraphics[width=0.525\textwidth]{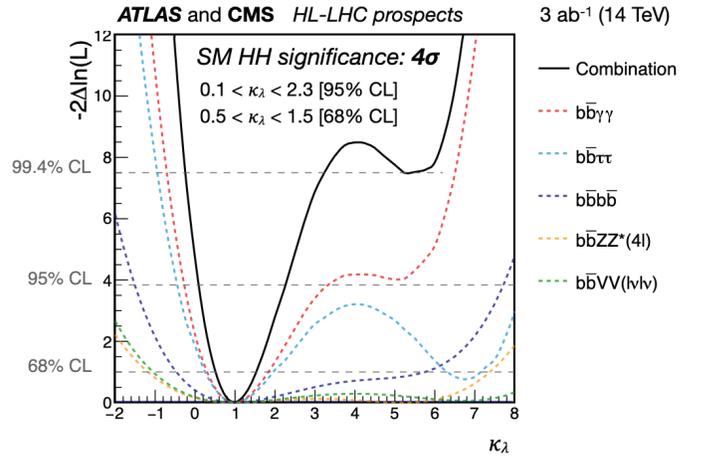} 
	\caption{Negative-log-likelihood scan as a function of $\kappa_{\lambda}$, calculated by performing a conditional signal$+$background
fit to the background and SM signal. The colored dashed lines correspond to the combined ATLAS and CMS results
by channel, and the black line their combination.  Taken from \cite{ATLAS:2022hsp}.\label{HLLHC2}}
\end{figure}

In July, 2022, Run 3 of the LHC commenced, with an expectation of collecting 250 fb$^{-1}$ of data over the next four years.
After completion of Run 3, preparations will begin for the High Luminosity phase of LHC (HL-LHC).  It is expected that the~HL-LHC will begin its run at the end of this decade, and will deliver 3 ab$^{-1}$ of data during its lifetime (after roughly ten years of operation).   The expected improvement of the experimentally measured Higgs boson properties is exhibited in Figure \ref{HLLHC1}.  

One of the most critical measurements not shown in Figure \ref{HLLHC1} is the triple Higgs coupling, which would provide the first evidence for the shape of the scalar Higgs potential away from its minimum.  The triple Higgs coupling $\lambda$ can be deduced from the observation of $HH$ production.  In particular, the $HH$ production amplitude consists of two contributions: one term that is proportional to $\lambda$ and a second term that is independent 
of~$\lambda$.  If the triple Higgs coupling is given by its SM value, then a combined analysis by the ATLAS and CMS Collaborations with the full HL-LHC dataset of 3 ab$^{-1}$ anticipates an observation of $HH$ production with a significance of $4\sigma$, as shown in Figure~\ref{HLLHC2}. Due to the destructive interference of the two contributions to the $HH$ production amplitude noted above, the triple Higgs coupling modifier 
$\kappa_\lambda$ (where $\kappa_\lambda=1$ corresponds to the SM expectation) is projected to be measured with an accuracy of $50\%$ at the HL-LHC, as indicated in Figure \ref{HLLHC2}.

Although the Higgs boson $H$ appears to be SM-like, there are numerous questions that must be addressed by future experiments.
Some of these questions can be addressed by the HL-LHC, whereas others will require a new generation of colliders.\footnote{The next generation of colliders under consideration includes $e^+ e^-$ colliders with a center of mass energy starting at 250 GeV and eventually reaching 1 TeV and beyond, a $\mu^+\mu^-$ collider with a center of mass energy between 3 and 10 TeV, and a $pp$ collider with a center of mass energy of 100 TeV~\cite{Narain:2022qud}.}  
The following Higgs wishlist consists of some key questions to be answered in future Higgs studies.
\vskip 0.1in

\begin{enumerate}
\item
What are the coupling strengths of $H$ to second generation quarks $(c,s)$?
\item
Can the coupling strengths of $H$ to first generation quarks $(u,d)$ and/or gluons be determined?
\item
Is it possible to measure the coupling strength of $H$ to electrons?
\item
Will sufficient precision ever exist to measure the invisible decay partial width expected in the SM ($H\to ZZ^*\to \nu\bar{\nu}\nu\bar{\nu}$)? How well can one constrain the branching ratio of $H$ into invisible final states (that might be present in certain BSM scenarios).
\item
 With what ultimate accuracy can one predict the properties (cross sections, partial widths, etc.) of the SM Higgs boson? What are the important missing theoretical computations that need to be done?
\item
 How well can one determine the Higgs self-coupling?
 To what extent (and with what accuracy) can one experimentally reconstruct the Higgs scalar potential?
 \item
 With what accuracy and reliability can one experimentally determine the total decay width of $H$?
 \item
 Will experimental deviations from SM Higgs boson properties, if observed, be convincing? Will they reveal a new energy threshold for BSM physics?
 \item
Do Higgs bosons couple to a dark sector made up of new particles that are completely neutral with respect to the SM (the so-called ``Higgs portal''\cite{Patt:2006fw})?
 \item
 Will convincing data emerge that implies that $H$ is not an elementary scalar but is in fact a composite state of more fundamental entities?
\end{enumerate}

\section{Beyond the SM Higgs boson}
\label{beyond}

One notable feature of the SM is the minimal structure of the scalar sector of the theory.  In light of the nonminimality of the fermion and gauge sectors, it is tempting to suggest that the scalar sector of the theory should be nonminimal as well.  Indeed, there are numerous motivations for considering an extended Higgs sector.   For example, extended Higgs sectors can provide a dark matter candidate and can modify the electroweak phase transition in a way that facilitates baryogenesis in the early universe.   In addition, many models of BSM physics (often introduced to address the naturalness problem of the Higgs sector) require the introduction of
additional scalar states beyond the SM Higgs boson.

One cannot arbitrarily extend the Higgs sector of the SM due to a number of important constraints arising from observed experimental data.  For example, the electroweak $\rho$ parameter, which is measured to be very close to 1~\cite{ParticleDataGroup:2022pth}, significantly restricts the possible scalar multiplets.
Given a scalar multiplet of weak isospin $T$ and hypercharge\footnote{The hypercharge $Y$ is normalized such that the electric charge of the scalar field is $Q=T_3+Y/2$.} $Y$,
\begin{equation} \label{di}
\rho\equiv \frac{m_W^2}{m_Z^2\cos^2\theta_W}=1\quad\Leftrightarrow\quad (2T+1)^2-3Y^2=1\,,
\end{equation}
independently of the scalar field vacuum expectation values.  The simplest solutions to the Diophantine equation given in equation~(\ref{di}) correspond to singlet scalars $(T,Y)=(0,0)$ and hypercharge-one complex scalar doublets $(T,Y)=(\tfrac12,1)$.

More generally, one can achieve a tree-level value of \mbox{$\rho=1$} by fine-tuning the parameters of the scalar potential if the following condition is satisfied~\cite{Gunion:1989we}:
\beq \label{finetune}
\sum_{T,Y}\bigl[4T(T+1)-3Y^2\bigr]|V_{T,Y}|^2 c_{T,Y}=0\,,
\eeq
where $V_{T,Y}\equiv\langle\Phi(T,Y)\rangle$ is the vacuum expectation value of the scalar field $\Phi$, and $c_{T,Y}=1$ [$c_{T,Y}=\half$] for a complex [real] representation of scalar fields.  For example, the Georgi-Machacek model~\cite{Georgi:1985nv}, which extends the SM Higgs sector by adding a hypercharge-two complex scalar triplet $(T,Y)=(1,2)$ and a hypercharge-zero real scalar triplet $(T,Y)=(1,0)$, satisfies the condition specified in equation~(\ref{finetune}) by choosing a scalar potential in which 
$V_{1,2}=V_{1,0}$.   However, this scalar potential is not stable with respect to radiative corrections, and thus requires a fine-tuning of scalar potential parameters to impose the required condition on the triplet scalar field vacuum expectation values.

 The experimental rarity of  flavor-changing neutral currents imposes significant restrictions on the structure of the Higgs couplings to fermion pairs.   
 This implies that Higgs-mediated flavor-changing neutral currents (FCNCs) must be significantly suppressed.   In addition, charged Higgs exchange at tree level (which can mediate $\overline{B}_d\to D^{(*)}\tau^{-}\bar{\nu}_\tau$) and at one loop (which can contribute to $b\to s\gamma$) yields significant constraints on the charged Higgs mass and the corresponding Yukawa couplings~\cite{Arbey:2017gmh}.   Extended Higgs sectors also provide new sources for CP violation, which are constrained by the nonobservation of electric dipole moments for the neutron and electron~\cite{Altmannshofer:2020shb}.
 Finally, given that the Higgs boson appears to be SM-like, any model that extends the Higgs sector beyond the SM must contain one scalar state that can be identified with the SM-like Higgs boson observed at the LHC.

The absence of tree-level scalar-mediated FCNCs and the presence of a SM-like Higgs boson in the scalar spectrum are features of certain limiting cases of extended Higgs models, called the flavor alignment limit and the Higgs alignment limit, respectively.
Flavor alignment~\cite{Pich:2009sp} posits that all Higgs-fermion Yukawa matrices are aligned with the corresponding fermion mass matrices (which eliminates flavor-changing neutral currents mediated by a neutral scalar exchange in the Higgs Lagrangian).  Higgs alignment~\cite{Craig:2013hca,Haber:2013mia,Asner:2013psa,Carena:2013ooa,BhupalDev:2014bir} posits that in a scalar field basis where only one of the Higgs doublet fields possesses a nonzero
vacuum expectation value, the associated neutral scalar field that resides in the same doublet is an approximate mass eigenstate field (to be identified with the SM-like Higgs boson observed at the LHC).

To illustrate the constraints arising from imposing these two alignment conditions, consider the two Higgs doublet model (2HDM), in which a second Higgs doublet is added to the SM~\cite{Branco:2011iw}.   The two Higgs doublet fields are denoted by $\Phi_1$ and $\Phi_2$.
After minimizing the scalar potential, the neutral components of the two doublet fields acquire vacuum expectation values $\langle\Phi_i^0\rangle=v_i/\sqrt{2}$, where \mbox{$v^2\equiv |v_1|^2+|v_2|^2\simeq (246~{\rm GeV})^2$.}  It is convenient to introduce the Higgs basis fields~\cite{Georgi:1978ri,Lavoura:1994yu,Lavoura:1994fv,Botella:1994cs,Branco:1999fs,Davidson:2005cw}:
\beq
H_1= v_1^*\Phi_1+v_2^*\Phi_2\,,\qquad\quad H_2=-v_2\Phi_1+v_1\Phi_2\,,
\eeq
which satisfy $\langle H^0_1\rangle=v/\sqrt{2}$ and $\langle H^0_2\rangle=0$.  The most general renormalizable gauge-invariant scalar potential, in terms of the Higgs basis fields, 
is given by~\cite{Davidson:2005cw}
 \beqa
 \mathcal{V}&=& Y_1 {H}_1^\dagger {H}_1+ Y_2 {H}_2^\dagger {H}_2 +[Y_3 
{H}_1^\dagger {H}_2+{\rm h.c.}] \nn \\[3pt]
&& +\,\half Z_1({H}_1^\dagger {H}_1)^2+\half Z_2({H}_2^\dagger {H}_2)^2
+Z_3({H}_1^\dagger {H}_1)({H}_2^\dagger {H}_2) \nn \\[3pt]
&& +Z_4({H}_1^\dagger {H}_2)({H}_2^\dagger {H}_1)
+\left\{\half Z_5({H}_1^\dagger {H}_2)^2 \right. \nn \\ 
&&\qquad \left.+\big[Z_6 {H}_1^\dagger
{H}_1 +Z_7  {H}_2^\dagger {H}_2\big] {H}_1^\dagger {H}_2+{\rm
h.c.}\right\}.
\eeqa
Minimization of the scalar potential fixes $Y_1=-\half Z_1 v^2$ and $Y_3=-\half Z_6 v^2$.
The charged Higgs squared mass is given by 
\beq
m_{H^\pm}^2=Y_2+\half Z_3 v^2\,.
\eeq
The neutral Higgs squared masses are given by the eigenvalues of a $3\times 3$ neutral Higgs squared mass matrix (see \cite{Haber:2006ue} for further details).

One can introduce a $\mathbb{Z}_2$ symmetry that enforces flavor alignment, which leads to the so-called Type I, II, X, and Y Higgs--fermion Yukawa interactions of the 2HDM~\cite{Hall:1981bc,Aoki:2009ha}.  If CP conservation is imposed on the scalar potential, then the Higgs basis field $H_2$ can be rephased such that all scalar potential parameters (and the vacuum expectation values $v_1$ and $v_2$) are real.
In this case, the neutral scalar spectrum consists of two CP-even scalars $h$ and $H$ (with $m_h\leq m_H$) and a CP-odd scalar $A$ with mass:
\beq
m_A^2=m_{H^\pm}^2+\half(Z_4-Z_5)v^2\,.
\eeq
The masses of $h$ and $H$ are obtained by diagonalizing the $2\times 2$ squared mass matrix:
\beq
\mathcal{M}_H^2=\begin{pmatrix} Z_1 v^2 & \quad Z_6 v^2 \\  Z_6 v^2 & \quad m_A^2+Z_5 v^2\end{pmatrix},
\eeq
with respect to the Higgs basis states,
$\{\sqrt{2}\,{\rm Re}~H^0_1-v$\,,\,$\sqrt{2}\,{\rm Re}~H^0_2\}$.  In particular,
\beq
\begin{pmatrix} H\\ h\end{pmatrix}=\begin{pmatrix} \cbma & -\sbma \\
\sbma & \phantom{-}\cbma\end{pmatrix}\,\begin{pmatrix} \sqrt{2}\,\,{\rm Re}~H_1^0-v \\ 
\sqrt{2}\,{\rm Re}~H_2^0
\end{pmatrix}\,,
\eeq
where $\alpha$ is the mixing angle of the matrix that diagonalizes the CP-even Higgs squared-mass matrix when written with respect to the basis of scalar fields, $\{\sqrt{2}\,{\rm Re}~\Phi_1^0-v_1\,,\,\sqrt{2}\,{\rm Re}~\Phi_2^0-v_2\}$, and $\tan\beta\equiv v_2/v_1$.

\begin{figure}[t!]
    \includegraphics[width=0.45\textwidth]{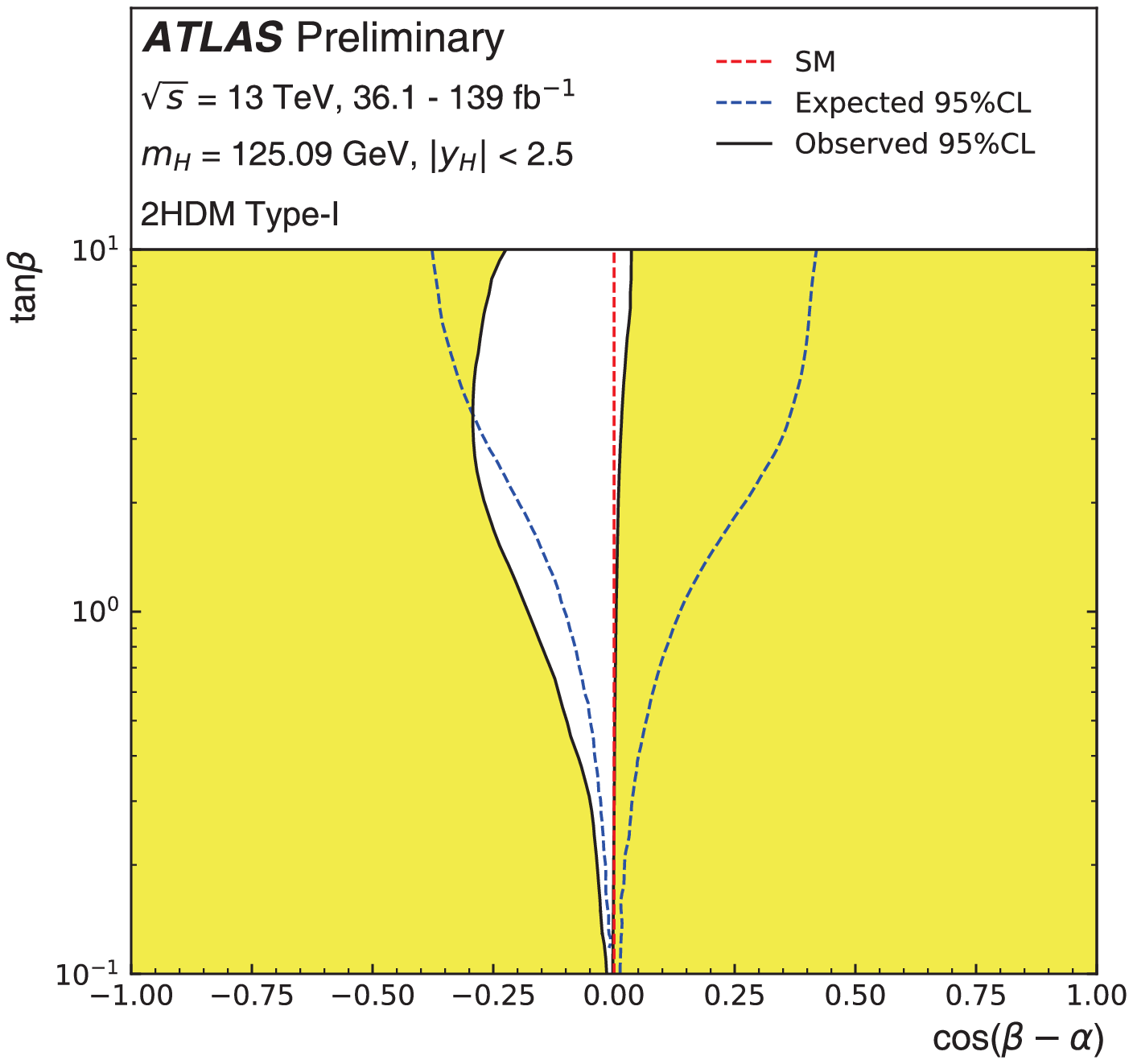} \\
    \includegraphics[width=0.45\textwidth]{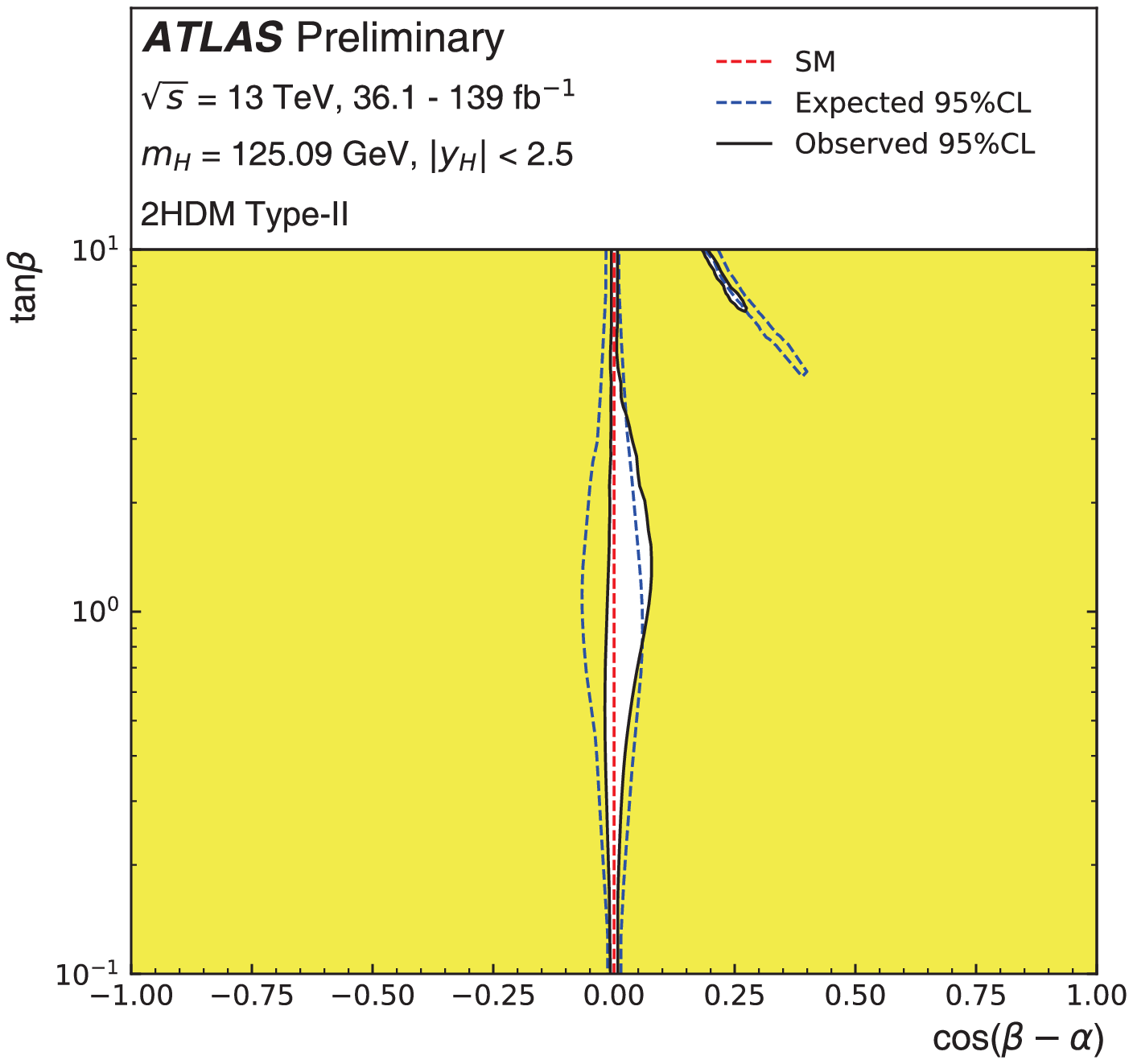}
	\caption{Regions excluded (at 95\% CL) by fits to the measured rates of the productions and decay of the Higgs boson (assumed to be $h$ of the 2HDM), highlighted in yellow. The observed best-fit values for $\cos(\beta-\alpha)$ are $-0.006$ for the Type-I 2HDM and $0.002$ for the Type-II 2HDM. Taken from \cite{ATLAS:2021vrm}.
}
	\label{types}
\end{figure}

The departure from the Higgs alignment limit is characterized by the parameter $\cos(\beta-\alpha)$, which is zero in the limit of exact Higgs alignment.  In the CP-conserving 2HDM in the approximate Higgs alignment limit, 
\begin{equation} \label{zeesix}
\cos(\beta-\alpha)\simeq \frac{Z_6 v^2}{m_{h}^2-m_{H}^2}\,, 
\end{equation}
where $h$ is the SM-like Higgs boson.   Thus, the Higgs alignment limit is approximately realized if either $m_{H}\gg m_{h}$ (called the decoupling limit~\cite{Haber:1989xc,Gunion:2002zf}) and/or if $|Z_6|\ll 1$.   In Figure \ref{types}, regions of the 2HDM parameter space highlighted in yellow are excluded by the current precision of the LHC Higgs data by the ATLAS Collaboration~\cite{ATLAS:2021vrm}, which implies that the parameters of the 2HDM must respect an approximate Higgs alignment limit in order to be consistent with current LHC observations.

One specialized version of the 2HDM of some interest is the inert doublet model (IDM)~\cite{Barbieri:2006dq,LopezHonorez:2006gr}, where $Y_3=Z_6=Z_7=0$.  This model possesses an exact $\mathbb{Z}_2$ symmetry, which remains unbroken after imposing the scalar potential minimum conditions.   This implies that the gauge fields and the Higgs basis field $H_1$
are even under the $\mathbb{Z}_2$ symmetry, whereas the Higgs basis field $H_2$ is odd under the $\mathbb{Z}_2$ symmetry.   
In particular, the IDM possesses an exact Higgs alignment in light of equation~(\ref{zeesix}).
The observed Higgs boson along with the neutral and charged Goldstone bosons reside completely inside the Higgs basis field $H_1$ with tree-level Higgs couplings that coincide with those of the SM Higgs boson.

Moreover, by insisting that all fermions are even under the $\mathbb{Z}_2$ symmetry, the resulting Yukawa interactions are of Type~I, thereby satisfying the requirement of flavor alignment.
The only fields of the model that are odd under $\mathbb{Z}_2$ are the charged Higgs boson pair $H^\pm$ and two new neutral Higgs bosons, which reside in the Higgs basis field $H_2$.   Assuming that the lightest $\mathbb{Z}_2$-odd particle (LOP) is neutral, then the LOP is stable and is a candidate for dark matter.  A parameter regime of the IDM exists in which the LOP accounts for the observed dark matter relic density, as shown in \cite{Goudelis:2013uca}.

More generically, if we allow for CP-violating scalar potential, then the physical scalars of the 2HDM consist of three neutral scalars $h_1$, $h_2$, and $h_3$ (with indefinite CP quantum numbers) and a charged scalar pair $H^\pm$.   The neutral scalars can be expressed in terms of the scalar doublet fields $\Phi_1$ and $\Phi_2$ and their vacuum expectation values,
$\langle{\Phi^0_i}\rangle=v\widehat{v}_i/\sqrt{2}$,
as follows:
\beq
h_k=\frac{1}{\sqrt{2}}\sum_{i=1}^2\left[({\overline\Phi\llsup{0}_i})^\dagger
(q_{k1} \widehat v_i+q_{k2}\widehat w_i)+{\rm h.c.}\right],
\eeq
where $\widehat{v}=(\cos\beta\,,\, e^{i\xi}\sin\beta)$, the (complex) unit vector 
$\widehat{w}$ is orthogonal to $\widehat{v}$,
and the shifted neutral fields are defined by 
$\overline\Phi_i\lsup{0}\equiv \Phi_i^0-\langle\Phi_i^0\rangle$. The quantities $q_{k1}$ and $q_{k2}$ (where $k\in\{1,2,3\}$) can be expressed in terms mixing angles that arise in the diagonalization of the $3\times 3$ neutral scalar squared mass matrix (detailed expressions can be found in \cite{Haber:2006ue}). 

In the general 2HDM, the scalar couplings to two vector bosons are given by
\beq
\mathscr{L}_{VVH}=\left(gm_W W_\mu^+ W^{\mu -}+\frac{2}{2\cos\theta_W} m_Z Z_\mu Z^\mu\right)\sum_k q_{k1}h_k\,.\label{VVH}
\eeq
It is also instructive to exhibit the Higgs-quark Yukawa couplings:\footnote{Equation~(\ref{Yukawas}) is easily extended to include the Higgs boson couplings to leptons.   Since neutrinos are massless in the two-Higgs doublet extension of the Standard Model, one obtains the Higgs boson couplings to charged leptons and neutrinos by making the following replacements: 
$D\to E=(e,\mu,\tau)$, $U\to N=(\nu_e,\nu_\mu,\nu_\tau)$, 
$\rho^D\to \rho^E$, $M_D\to M_E$, and $M_U\to M_N$, where $M_E$ is the diagonal charged lepton mass matrix and $M_N=0$.}
\beqa
 \mathscr{L}_Y&= &-\frac{1}{v}\sum_k \overline U \biggl\{ q_{k1}M_U +\frac{v}{\sqrt{2}}\left[
q^*_{k2}\,\rho^U P_R+
q_{k2}\,\rho^{U\dagger} P_L\right]\biggr\}U h_k
 \nonumber \\
&& - \frac{1}{v}\sum_k\overline D
\biggl\{ q_{k1}M_D+ \frac{v}{\sqrt{2}}\left[
q_{k2}\,\rho^{D\dagger} P_R+
q^*_{k2}\,\rho^D P_L\right]\biggr\}Dh_k
\nonumber \\
&& -\biggl\{\overline U\left[K\rho^{D\dagger}
P_R-\rho^{U\dagger} KP_L\right] DH^+
+{\rm
h.c.}\biggr\}\,, \label{Yukawas}
\eeqa
where $P_{R,L}\equiv \tfrac12(1\pm\gamma_5)$, $U=(u,c,t)$ and $D=(d,s,b)$ are three families of quark fields, with corresponding \textit{diagonal} up-type and down-type quark mass matrices $M_U$ and $M_D$, respectively, $K$ is the CKM mixing matrix and $\rho^U$, $\rho^D$ are complex $3\times 3$ matrices.

In the exact Higgs alignment limit,
\beq
q_{11}=q_{22}=-iq_{32}=1\quad \text{and}\quad q_{21}=q_{31}=q_{12}=0\,.
\eeq
Inserting these values into equations~(\ref{VVH}) and (\ref{Yukawas}), it follows that in the exact Higgs alignment limit, the properties of $h_1$ coincide with those of the SM Higgs boson.
However, since the matrices $\rho^U$, $\rho^D$, and $\rho^E$ are in general nondiagonal (complex) matrices, it follows that tree-level FCNCs mediated by $h_2$ and $h_3$ are present, along with new CP-violating phases in the $h_2$, $h_3$, and $H^\pm$ couplings to fermions.

\begin{figure}[t]
\centering
    \includegraphics[width=0.46\textwidth]{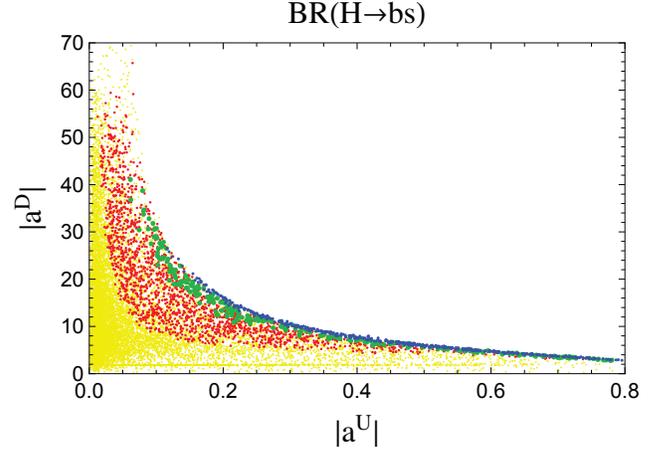}  
 	\caption{Branching ratio (BR) of the CP-even Higgs boson $H\to \bar{b}s + b\bar{s}$ for $m_H=400$~GeV
obtained by scanning the parameter space, with $\cos(\beta-\alpha)=0$ and $|a^E|=10$ fixed, as a function of the flavor alignment parameters $a^U$ and $a^D$.
	The yellow, red, green, and blue points correspond to BR$(H \to \bar{b}s + b\bar{s}) < 0.0005$, [0.0005, 0.01], [0.01, 0.1], and $> 0.1$, respectively.
 Taken from \cite{Gori:2017qwg}.
	}
	\label{bs}
\end{figure}

To eliminate flavor-nondiagonal and CP-violating Yukawa couplings of $h_2$ and $h_3$ to quarks and leptons, one can impose flavor alignment by assuming that
\beq \label{falign}
\rho^F=a^F M_F\,,\qquad \text{for $F=U,D,E$}\,,
\eeq
where the quantities $a^F$ are real flavor alignment parameters.  Equation (\ref{falign}), if implemented generically, is not stable under renormalization group evolution~\cite{Ferreira:2010xe}.  The diagonality condition can be imposed either at the electroweak scale by fine-tuning~\cite{Pich:2009sp} 
or at an energy scale $\Lambda\gg v$ where equation~(\ref{falign}) is satisfied as a consequence of some new ultraviolet physics
(e.g., see \cite{Serodio:2011hg,Egana-Ugrinovic:2019dqu}), in which case tree-level scalar-mediated FCNCs are generated at the electroweak scale and provide potential signals for discovery~\cite{Gori:2017qwg}.  For example, a nonzero branching ratio for $H\to \bar{b}s+b\bar{s}$ can be generated, whose value depends on the values of $a^U$ and $a^D$, as shown in Figure \ref{bs}.

Alternatively, as previously noted, one can impose a $\mathbb{Z}_2$ discrete symmetry on the Higgs Lagrangian to eliminate tree-level scalar-mediated FCNCs without the fine-tuning of parameters.  Different choices of the discrete symmetries yield the Types I, II, X, and Y Yukawa couplings previously mentioned.  
Such models are special cases of equation (\ref{falign}).  For example, Type I corresponds to $a^D=a^E=a^U$ and Type II corresponds to $a^D=a^E=-1/a^U$.

\begin{figure}[t!]
\centering
    \includegraphics[width=0.45\textwidth]{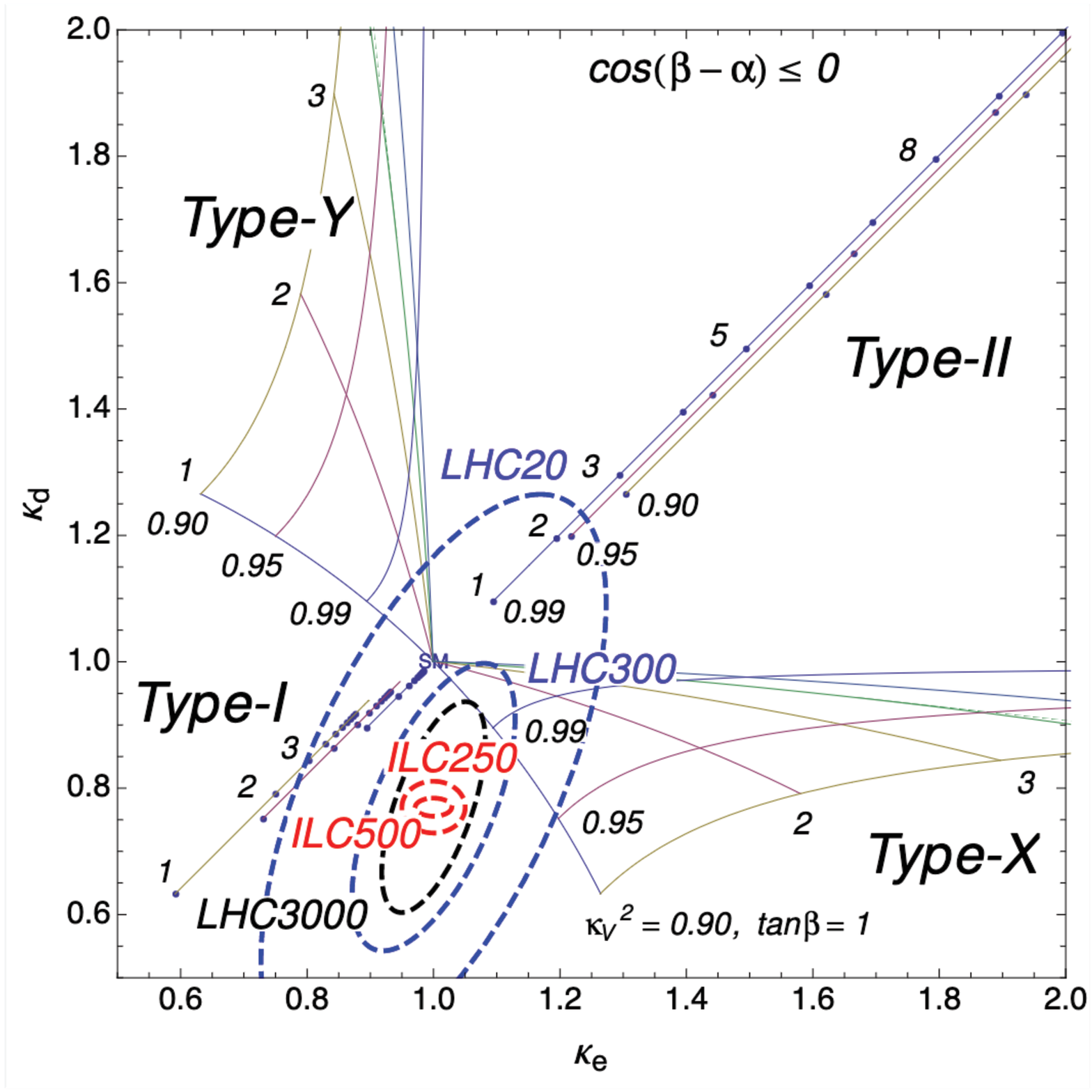}  \\
     \includegraphics[width=0.45\textwidth]{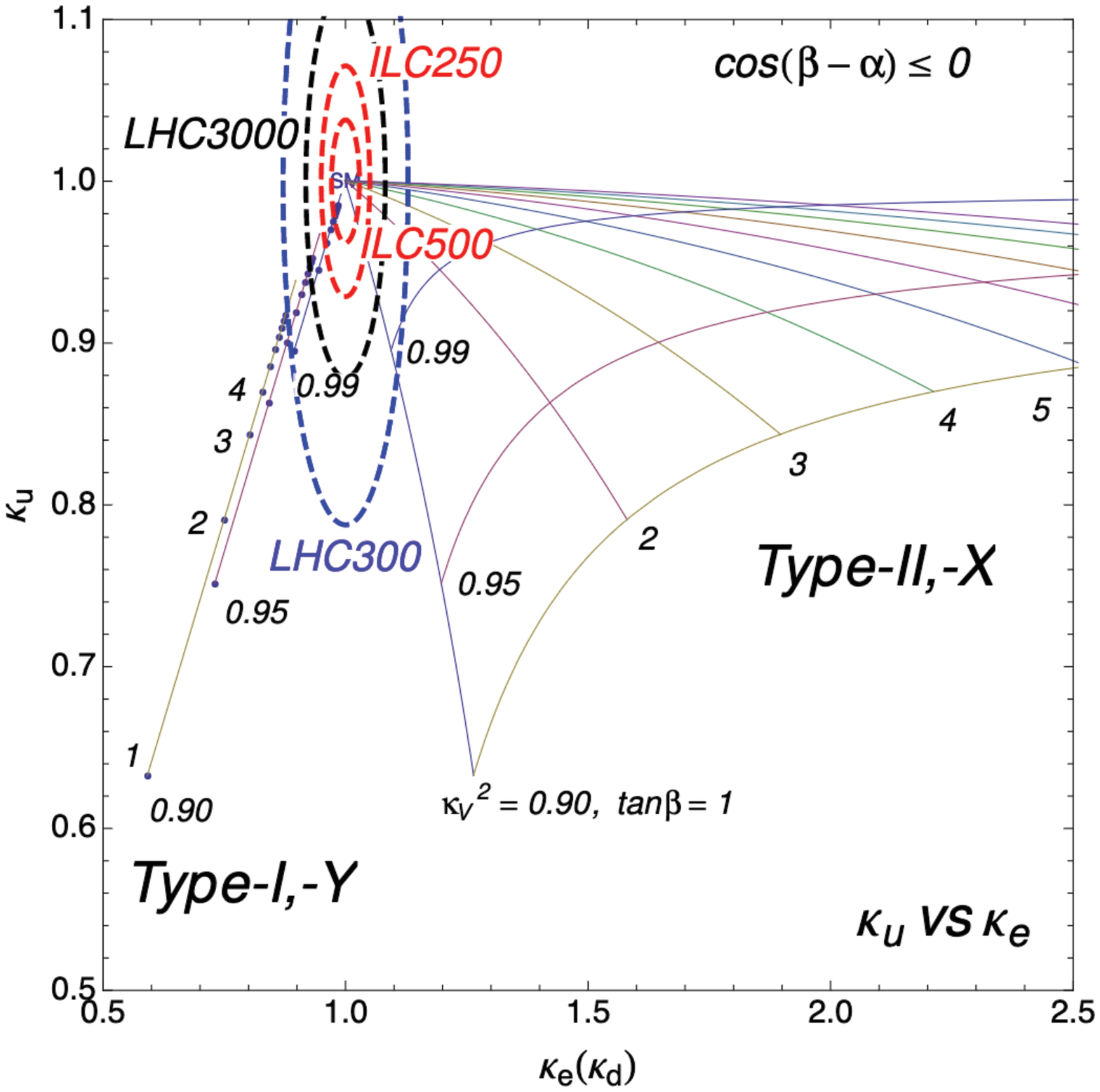} 
	\caption{The scaling factors for the Yukawa interaction of the SM-like Higgs boson in 2HDMs in the case
of $\cos(\beta-\alpha)<0$.  Taken from \cite{Kanemura:2014bqa}.
	}
	\label{fingerprint}
\end{figure}

As noted in Section~\ref{status},
the properties of the Higgs boson discovered at the LHC approximate those of the SM Higgs boson. 
Consequently, if additional Higgs bosons exist, then the Higgs alignment limit of the extended Higgs sector must be approximately realized. 
Although no additional 
scalar states beyond the Higgs boson with a mass of 125 GeV have been found so far at the LHC, evidence for additional scalar states (which would arise in models of extended Higgs sectors) and/or other BSM phenomena could emerge in future runs at the LHC or at new collider facilities that might be constructed in the future.  The discovery of any additional scalar states would
yield important clues to the underlying structure of the nonminimal Higgs sector.  As an example, the authors of \cite{Kanemura:2014bqa} provided some techniques for discriminating among different choices for the Yukawa interactions of the 2HDM, as illustrated in Figure \ref{fingerprint}.  

However, a more flexible framework for analyzing the properties of the new scalars should avoid imposing model assumptions that are stronger than those required for the consistency with experimental constraints.   Instead of imposing the Types I, II, X, and~Y frameworks on the 2HDM Yukawa couplings, one could instead impose a weaker flavor alignment constraint that is determined phenomenologically~\cite{Pich:2009sp} (instead of being a consequence of a symmetry).  For example,
\cite{Connell:2023jqq} provides an example of a scenario where new scalars of a flavor-aligned 2HDM are discovered that cannot be described by Yukawa couplings of Types I, II, X, and~Y.   If evidence for an extended Higgs sector is revealed, then one should rely on the experimental observations (with minimal theoretical prejudice) to determine the structure of the corresponding Yukawa couplings.

The Higgs sector can serve as a portal to BSM physics.   Indeed, many models of BSM physics require an extension of the SM Higgs sector.  For example, the Higgs sector of the minimal supersymmetric extension of the SM (MSSM) is a 2HDM with Type II Yukawa couplings and a CP-conserving scalar potential whose dimensionless couplings are constrained by supersymmetry.  The supersymmetry-breaking squared mass terms of the scalar potential are responsible for electroweak symmetry breaking.   The MSSM possesses a SM-like Higgs boson $h$ in the decoupling limit (where $m_A\gg m_h$).

\begin{figure}[t!]
\centering
    \includegraphics[width=0.45\textwidth]{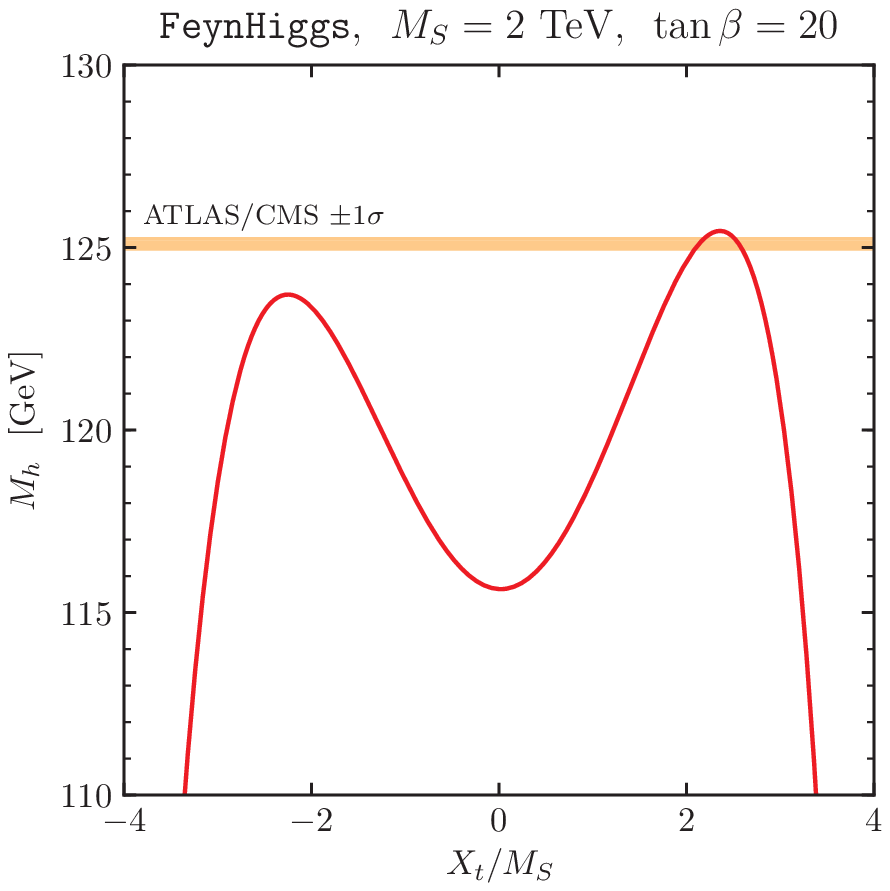}  \\
     \includegraphics[width=0.45\textwidth]{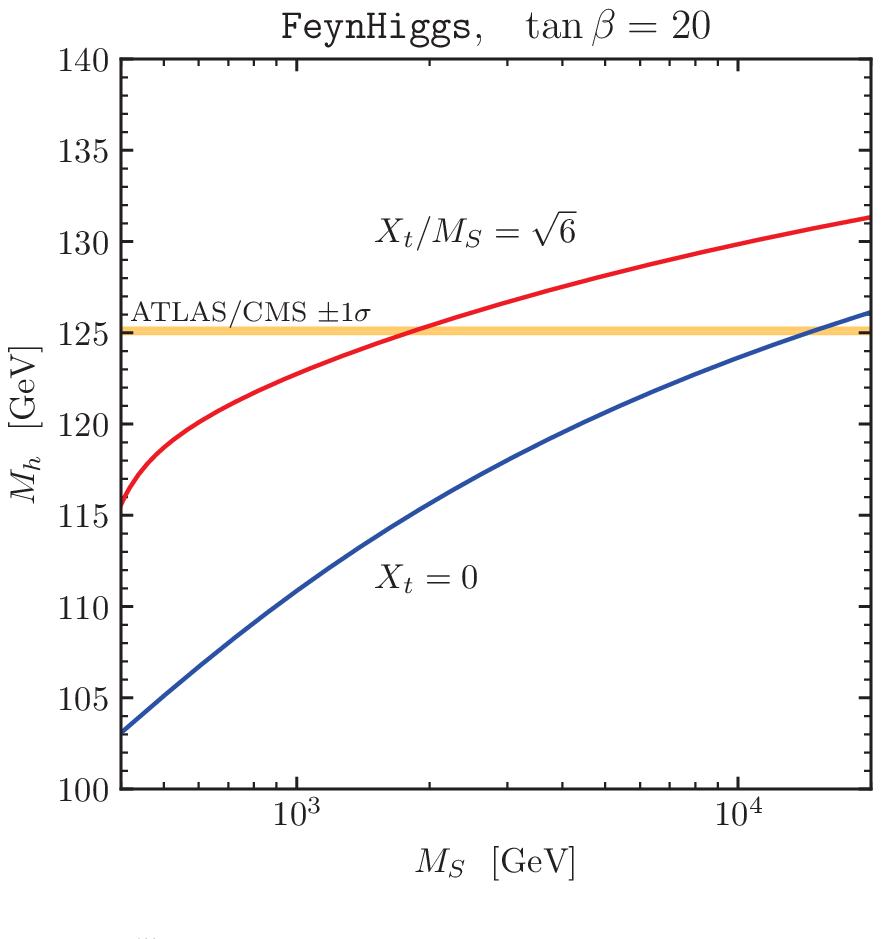} 
	\caption{The lighter CP-even Higgs mass in the MSSM as a function of a common supersymmetric mass parameter $M_S$ and of the stop mixing parameter $X_t$ (normalized to $M_S$).  Taken from \cite{Slavich:2020zjv}.
	}
	\label{MSSMhiggsmass}
\end{figure}

At the tree level, the MSSM yields $m_h\leq m_Z$ in conflict with observation.  However, radiative corrections raise the upper limit; an approximate one-loop computation produces the following result:
\beq
m_{h}^2\simeq  m_{Z}^2\cos^2 2\beta+{\displaystyle\frac{3g^2 m_{t}^4}{8\pi^2 m_{W}^2}}\left[\ln\left({\displaystyle\frac{\msusyy}{m_{t}^2}}\right)+
{\displaystyle\frac{|X_t|^2}{\msusyy}}
\left(1-{\displaystyle\frac{|X_t|^2}{12\msusyy}}\right)\right],
\eeq
where $X_t$ is the top squark (stop) mixing parameter and $M_S$ is the geometric mean of the two top squark masses (further details can be found in \cite{Dreiner:2023yus}).  The results of a more complete computation (which includes higher-order loop calculations and renormalization group improvements) are shown in Figure~\ref{MSSMhiggsmass}.  These results suggest that  supersymmetric particle masses (characterized by $M_S$) are likely to be closer to 10 TeV (rather than 1 TeV as initially anticipated), thereby explaining why no supersymmetric particles have yet been observed at the LHC.

Concerning Higgs physics beyond the SM, a second Higgs wishlist can be constructed that poses additional questions that one hopes can be answered in future Higgs studies.
\vskip 0.1in

\begin{enumerate}
\item
How many generations of Higgs scalars exist at or below the TeV scale and what are their electroweak quantum numbers?  
\item
Are there any new elementary scalars not yet discovered with masses below the mass of the SM-like Higgs boson?  For example, do axion-like particles exist?
\item
How small is the departure from the Higgs alignment limit, and what is the underlying mechanism that yields an approximate Higgs alignment?
\item
Does the unitarization of the scattering of longitudinal gauge bosons require additional scalars from an extended scalar sector?
\item
How small is the departure from the flavor alignment limit of the neutral Higgs--fermion Yukawa couplings?
Will quark/lepton flavor off-diagonal couplings of neutral scalars be observed? 
\item
Are there new sources of CP violation associated with the extended scalar sector? Can these be experimentally observed (and can the corresponding sources be identified)?
\item
How does the extended scalar sector affect the electroweak phase transition? Does it permit electroweak baryogenesis? Does it play other significant roles in early universe cosmology (e.g., inflation)?  Will future gravitational wave experiments shed any light on these matters?
\item
If additional scalars are discovered, how will these discoveries impact the question of the stability of the electroweak vacuum?
\item
Do neutral (inert) scalars comprise a significant fraction of the dark matter?
\item
How does the scalar sector inform the identification of the BSM physics?
Will this shine any light on the apparent unnatural nature of the SM Higgs sector?
\end{enumerate}

\section{Conclusions}

Future Higgs studies have the potential to address many profound questions concerning the theory of fundamental particles and their interactions.   With expectations of data samples that are 20 times larger than the Higgs data collected in Runs 1 and 2 of the LHC, there are considerable opportunities for discoveries at the LHC during the next 10 to 15 years.

Nevertheless, Nature may demand more precision than the LHC can ultimately provide.  As the particle physics community ponders the roadmap of future collider facilities, one should consider which paths provide the best opportunities for addressing the questions posed by the two Higgs wishlists presented in Sections~\ref{status} and \ref{beyond}.

\section*{Acknowledgments}

An earlier version of this review, which was originally based on the summary talk given at the 2021 SLAC Summer Institute, 
was contributed to the CERN celebration of the 10th anniversary of the discovery of the Higgs boson.  I am grateful to Tom Rizzo for
the original SLAC invitation and to Panagiotis Charitos for the opportunity to participate in the CERN celebration.
I would also like to thank Stefan Pokorski for inviting me to present this review at the PLANCK 2023 meeting in Warsaw, Poland.
In addition, I am pleased to acknowledge the
inspiring working atmosphere of the Aspen Center for
Physics, supported by the National Science Foundation
Grant No. PHY-1066293, where this write-up was prepared.   H.E.H. is supported in
part by the U.S. Department of Energy Grant No. DE-
SC0010107. 



\end{document}